\documentclass[12pt,eadjoint tfnpsf]{article}

\usepackage{amsbsy,amssymb,latexsym,amsfonts}
\usepackage{graphicx}
\usepackage{amsbsy,amssymb,latexsym,amsfonts}
\usepackage{amsmath,epsfig}
\usepackage{bm}
\usepackage{graphicx}

\RequirePackage[dvips,usenames]{color}
\definecolor{fireblick}{rgb}{0.698039,0.133333,0.133333}

\newcommand{\beq}{\begin{equation}}
\newcommand{\eeq}{\end{equation}}
\newcommand{\bea}{\begin{eqnarray}}
\newcommand{\eea}{\end{eqnarray}}

\newcommand{\CO}{{\mathcal O}}

\newcommand\tr{\mathrm{tr}}

\numberwithin{equation}{section}
\newcommand{\bel}[1]{\begin{equation}\label{#1}}                     
\newcommand{\bal}[1]{\begin{eqnarray}\label{#1}}                     
\newcommand{\be}{\begin{equation}}
\newcommand{\ee}{\end{equation}}
\def\beq{\begin{equation}}
\def\eeq{\end{equation}}

\newcommand{\mat}[1]{\begin{pmatrix} #1 \end{pmatrix}}
\renewcommand{\thefootnote}{\fnsymbol{footnote}}

\def\l{\lambda}

\def\d{\delta}

\def\G{\Gamma}

\def\CO{{\mathcal O}}

\def\12{\frac{1}{2}}

\def\gh{\mathrm{ghost}}
\def\gf{\mathrm{gf}}

\def\tr{\mathrm{tr}}

\def\vt{\tilde{v}}

\def\lt{\tilde{\l}}

\begin{document}
%%%%%%%%%%%%%%%%%%%%%%%%%%%%%%%%%%%%%%%%%%%%%%%%%%%%%%%%%%%%%%%%%%%%%%%%%%%%%%%%%%%%%%%%%%
%
% title page
%
%%%%%%%%%%%%%%%%%%%%%%%%%%%%%%%%%%%%%%%%%%%%%%%%%%%%%%%%%%%%%%%%%%%%%%%%%%%%%%%%%%%%%%%%%%
\begin{titlepage}

%%%%%%%%%%%%%%%%%%%% preprint # %%%%%%%%%%%%%%%%%%
\begin{flushright}
\normalsize
%\filename
~~~~
September, 2010 \\
OCU-PHYS 337 \\
\end{flushright}
%%%%%%%%%%%%%%%%%%%%%%%%%%%%%%%%%%%%%%%%%%%%%%%%%%

\bigskip
\bigskip

%%%%%%%%%%%%%%%%%%%% title %%%%%%%%%%%%%%%%%%%%%%%
\begin{center}
{\Large\bf   
Effects of Matrix Orientifolding to Two-Loop Effective Action 
 of Bosonic IIB Matrix Model}
\end{center}
%%%%%%%%%%%%%%%%%%%%%%%%%%%%%%%%%%%%%%%%%%%%%%%%%%

\bigskip

%%%%%%%%%%%%%%%%%%% authors %%%%%%%%%%%%%%%%%%%%%%
\begin{center}
{%
R. Yoshioka$^a$\footnote{e-mail: yoshioka@sci.osaka-cu.ac.jp}
}
\end{center}
%%%%%%%%%%%%%%%%%%%%%%%%%%%%%%%%%%%%%%%%%%%%%%%%%%

\bigskip

%%%%%%%%%%%%%%%%%%% affiliation %%%%%%%%%%%%%%%%%%%
\begin{center}
%$^a$ \it Department of Mathematics and Physics,
%Graduate School of Science\\
%Osaka City University\\

\medskip

$^a$ \it Osaka City University Advanced Mathematical Institute (OCAMI)

\bigskip

3-3-138, Sugimoto, Sumiyoshi, Osaka, 558-8585, Japan \\

\end{center}

%%%%%%%%%%%%%%%%%%%%%%%%%%%%%%%%%%%%%%%%%%%%%%%%%%%

\bigskip

%%%%%%%%%%%%%%%%%%%% abstract %%%%%%%%%%%%%%%%%%%%%
\begin{abstract}
We study the spacetime structures which are described by the IIB matrix model 
 with orientifolding. 
Matrix orientifolding that preserves supersymmetries yields the mirror image point  
 with respect to a four-dimensional plane
 for each spacetime point that corresponds to the eigenvalue of the bosonic matrix.
In order to consider the upper bound on the distance between two eigenvalues 
 in this model, we calculate the effective action for the eigenvalues up to two-loop.  
The eigenvalues distribute in a tubular region 
 around the four-dimensional plane. 
\end{abstract}
%%%%%%%%%%%%%%%%%%%%%%%%%%%%%%%%%%%%%%%%%%%%%%%%%%%

\vfill

\setcounter{footnote}{0}
\renewcommand{\thefootnote}{\arabic{footnote}}

\end{titlepage}

%%%%%%%%%%%%%%%%%%%%%%%%%%%%%%%%%%%%%%%%%%%%%%%%%%%%%%%%%%%%%%%%%%%%%%%%%%%%%%%%%%%%%%%%%
%%%%%%%%%%%%%%%%%%%%%%%%%%%%%%%%%%%%%%%%%%%%%%%%%%%%%%%%%%%%%%%%%%%%%%%%%%%%%%%%%%%%%%%%%
\section{Introduction}
%%%%%%%%%%%%%%%%%%%%%%%%%%%%%%%%%%%%%%%%%%%%%%%%%%%%%%%%%%%%%%%%%%%%%%%%%%%%%%%%%%%%%%%%%

%matrix model and  orientifolding... 
The reduced matrix models are proposed to enable nonperturbative studies of 
 strings {\cite{BFSS}-\cite{Tay}}.
They are obtained from Yang-Mills theory by the dimensional reduction{\cite{EK}}. 
The eigenvalues of the matrices represent spacetime points, while the remaining degrees 
 of freedom mediate interactions between the spacetime points. 
The effective dynamics of the spacetime points is obtained by carrying out the integrations 
of the off-diagonal elements.
The formation of our spacetime have been variously attempted. 
For example, branched polymers \cite{AIKKT}, 
generalied monopoles \cite{CIK2, IM}, 
orbifolding \cite{AIS,IY} and more have been used. 
The spontaneous breakdown of Lorentz symmetry to lower dimension 
 due to fermion determinant is also developed \cite{ANO}.

The USp matrix model is introduced as the orientifolding 
 of the IIB matrix model, which preserves the maximal supersymmetry \cite{IToku,ITsu}. 
In \cite{IY2}, we have seen that there is a long distance attraction between 
 the spacetime points up to the one-loop corrections in this model. 
Moreover it was found that two-body force in the short distance is repulsive 
  by calculations at the model with lower rank matrices, $usp(2)$ and $usp(4)$. 

We continued to study the USp matrix model, 
\be
 S = - \frac{1}{4g^2} \tr [v_{M} , v_{N}]^2 - \frac{1}{2g^2} \tr \Psi \G^M [v_M, \Psi], 
\ee 
where $M=0, 1, \cdots, 9$ and $\Psi$ is a ten-dimensional Majorana-Weyl spinor. 
The matrices $v_M$ take the following form: 
\begin{eqnarray}
 v_{\mu} = \mat{M_{\mu} & N_{\mu} \\ N^*_{\mu} & - M^t_{\mu}}, ~~~~~~ %&~~~~& 
 %\Psi_{\a} = \mat{O_{\mu} & P_{\mu} \\ P^*_{\mu} & - O^t_{\mu}}, 
 %\cr
 v_{m} = \mat{A_{m} & B_{m} \\ -B^*_{m} & A^t_{m}}, %&~~~~&
 %\Psi_{\a'} = \mat{C_{\a'} & D_{\a'} \\ -C^*_{\a'} & D^t_{\a'}}
\end{eqnarray}
where $\mu = 0,1,2,3,4,7$ and $m = 5,6,8,9$ and 
 $M_{\mu}$ and $A_{m}$ are $N \times N$ Hermitian matrices 
  and $N_{\mu}$($B_{m}$) are $N \times N$ (anti-)symmetric. 
Selecting one of these two representations for each of the matrix coordinates 
 is referred to as matrix orientifolding in this paper.
In what follows suppose that the matrices labeled by Greek letters $\mu,\nu,\cdots$ belong 
 to defining representation (we also call this as adjoint reprensantation) 
 of $usp$ Lie algebra and 
 those labeled by Roman letters $m,n,\cdots$ belong to antisymmetric representation. 
This splitting of the representation has taken place 
 in order to the preserve $8+8$ supersymmetries after the orientifolding 
 of the IIB matrix model with $16+16$ supersymmetries and is almost a unique way to choose. 
By construction, the ten-dimensional Lorentz covariance is broken 
 to four- and six-dimensional ones explicitly. 
The upshot is that matrix orientifolding inevitably introduces 
 spacetime directional asymmetry.
In this paper we restrict, however, our attention to the bosonic part.

We consider the effective dynamics for the 
 eigenvalues of the bosonic matrices,
 which are obtained by the integrations of the offdiagonal elements. 
In \cite{HNT}, the author found the upper bound on the extent of spacetime in the 
 bosonic IIB matrix model by calculating the two-loop corrections 
 to the effective action for the eigenvalues. 
By using the similar prescription, 
 we discuss what feature the spacetime described by the USp matrix model has. 

The content of this paper is as follows:
In the next section we calculate concretely the two-loop effective action for the eigenvalues. 
In section three the spacetime constituted by the USp matrix model is discussed.

%%%%%%%%%%%%%%%%%%%%%%%%%%%%%%%%%%%%%%%%%%%%%%%%%%%%%%%%%%%%%%%%%%%%%%%%%%%%%%%%%%%%%%
%%%%%%%%%%%%%%%%%%%%%%%%%%%%%%%%%%%%%%%%%%%%%%%%%%%%%%%%%%%%%%%%%%%%%%%%%%%%%%%%%%%%%%
\section{Two-loop corrections of the bosonic USp matrix model}
%%%%%%%%%%%%%%%%%%%%%%%%%%%%%%%%%%%%%%%%%%%%%%%%%%%%%%%%%%%%%%%%%%%%%%%%%%%%%%%%%%%%%%
In this section, we consider the bosonic part of the USp matrix model.
%Because the restriction for the splitting of the representation is due to supersymmetry,
Then we can arbitrarily choose the number of the bosonic coordinates belonging to 
 either adjoint representation or antisymmetric representation, 
 because this restriction is due to supersymmetry. %as mentioned in Introduction. 
We denote the number of the directions of the adjoint and antisymmetric representation 
by $D_{ad}$ and $D_{as}$, respectively.
The action is 
\be
 S_b = - \frac{1}{4g^2} \tr [v_{M} , v_{N}]^2, 
 \label{action:b}
\ee 
We decompose the matrices $v_{M}$ into the diagonal and the off-diagonal parts, 
\begin{align}
 v_M = x_M + \vt_M. %, ~~~ \Psi_A = \eta_A + \Pst_A. 
\end{align} 
The diagonal parts $x_M$ %of the matrices belonging to the adjoint and antisymmetric representation 
 are respectively given by 
\be
 x_{\mu} = \mat{X_{\mu}&0\\ 0&-X_{\mu}}, ~~~ x_{m} = \mat{X_{m}&0\\0&X_{m}}, 
 \label{diag_ele}
\ee
where 
\be
 X_M = \mat{x_M^1&&&\\&x_M^2&&\\&&\ddots&\\&&&x_M^N}. 
\ee 
Since the eigenvalues of the matrix correspond to the spacetime points, 
 the diagonal matrices (\ref{diag_ele}) represent the spacetime 
 configuration as shown in Figure 1. 
Note that the lower half of the diagonal elements is not independent variables 
 and correspond to the mirror image points with respect to the $D_{as}$-dimensional plane 
 spanned by the directions of antisymmetric representation. 
The upper half of the matrices thus represents spacetime.

%--------------------------FIGURE-----------------------------
\begin{figure}[h]
\begin{center}
\includegraphics[scale=0.6]{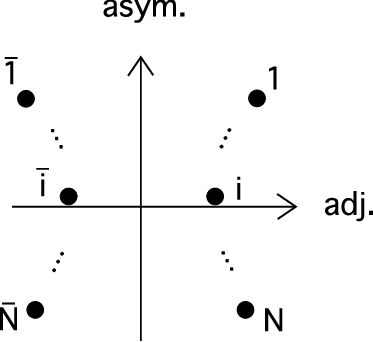}
\caption{
Each node represents the diagonal elements of the matrices $x_M$, 
 which are regarded as spacetime points and their images in ten dimensional spacetime.
 Because  ten $2N \times 2N$  bosonic matrices appear in the USp matrix model, 
 there are 2$N$ points in this figure.  
 Clearly, the spacetime points and their images are symmetric with respect to the 
 four-dimensional plane spanned by the antisymmetric directions. 
}
%\label{fig:diag}
\end{center}
\end{figure}
%--------------------------FIGURE-----------------------------

Using the eigenvalues $\l_M$ of $v_M$ and the off-diagonal elements $\vt_M$, 
 we can rewrite the diagonal elements $x_M$ as 
\begin{align}
 x_{M}^{i'} &\sim \l_{M}^{i'} 
   - \sum_{j' \neq i'} \frac{\vt_{M}^{i'j'} \vt_{M}^{j'i'}}{\l_{M}^{i'} - \l_{M}^{j'}} 
 	\equiv \l_{M}^{i'} + x'_M, 
\end{align} 
up to $O(\vt^2)$, where $i' = 1, 2, \cdots, 2N$. 
The difference $x'_M$ between the diagonal element and eigenvalue becomes important for
 higher loop corrections though it can be neglected for one-loop.

The quadratic action for the off-diagonal elements is 
\begin{align}
 S' &= S_b^{(2)} + S_{\gf}^{(2)} + S_{\gh}^{(2)} \cr 
    &= \frac{1}{g^2} \sum_{i,j} \left[
   	\{ (\l_{\mu}^{ij})^2 + (\l_{m}^{ij})^2 \} M_{\nu}^{ij} M^{\nu ij*} 
           + \{ (\lt_{\mu}^{ij})^2 + (\l_{m}^{ij})^2 \} N_{\nu}^{ij} N^{\nu ij*} 
        \right. \cr 
        &\hspace{2cm} \left.
           + \{ (\l_{\mu}^{ij})^2 + (\l_{m}^{ij})^2 \} A_{n}^{ij} A^{n ij*}
           + \{ (\lt_{\mu}^{ij})^2 + (\l_{m}^{ij})^2 \} B_{n}^{ij} B^{n ij*}
        \right] \cr 
    &~~~~ - \frac{1}{g^2} \tr [\l_{M},b][\l_{M},c], 
\label{action:quad}
\end{align}
where 
\be
 \l^{ij}_{M} = \l^i_{M} - \l^j_{M}, ~~~~ \lt^{ij}_{M} = \l^i_{M} + \l^j_{M}. 
\ee 
Here we have added the following gauge fixing term and the ghost term to the action: 
\begin{align}
S_{\gf} &= - \frac{1}{2g^2} \tr[\l_{M} , \vt_M ], ~~~
S_{\gh} = - \frac{1}{g^2} \tr[\l_{M} , b][v_{M} , c]. 
\end{align}
In Eq.(\ref{action:quad}), the quadratic parts $S_{\gf}^{(2)}$ and $S_{\gh}^{(2)}$
  have been included. 
The ghost $c$ and anti-ghost $b$ belong to adjoint representation, 
\be
 c = \mat{c_{(1)} & c_{(2)} \\ c_{(2)}^* & -c_{(1)}^t}, ~~~~ 
 b = \mat{b_{(1)} & b_{(2)} \\ b_{(2)}^* & -b_{(1)}^t}, 
\ee 
where $b_{(1)}$ and $c_{(1)}$ are Hermitian and $b_{(2)}$ and $c_{(2)}$ are symmetric. 
The propagators can be read off from the quadratic action (\ref{action:quad}) as follows: 
\begin{align}
 &\langle A_{M}^{*ij} A_{N}^{kl} \rangle  
  = g^2 \frac{1}{(\l^{ij})^2} \d^{ik} \d^{jl} \d_{MN}, \cr 
 &\langle N_{\mu}^{*ij} N_{\nu}^{kl} \rangle  
  = g^2 \frac{1}{(\lt^{ij})^2} \frac{1}{2}(\d^{ik} \d^{jl} 
  + \d^{il} \d^{jk}) \d_{\mu\nu}, \cr  
 &\langle B_{m}^{*ij} B_{n}^{kl} \rangle  
  = g^2 \frac{1}{(\lt^{ij})^2} \frac{1}{2}(\d^{ik} \d^{jl} 
  - \d^{il} \d^{jk}) \d_{\mu\nu}, \cr 
 &\langle c_{(1)}^{ij} b_{(1)}^{kl} \rangle 
  = g^2 \frac{1}{(\l^{ij})^2} \d^{il} \d^{jk}, \cr
 &\langle c_{(2)}^{ij} b_{(2)}^{kl} \rangle 
  = g^2 \frac{1}{(\lt^{ij})^2} \frac{1}{2}(\d^{ik} \d^{jl} + \d^{il} \d^{jk}), 
\end{align}
where $\langle ~ \rangle$ represents expectation value defined by 
\be
 \langle \CO \rangle = \frac{\int \CO e^{-S'}}{\int e^{-S'}}, 
\ee
and $A_M = (M_{\mu},A_{m})$ whose eigenvalues are ten-dimensional spacetime coordinates 
 and $(\l^{ij})^2 = (\l^{i}_{M} - \l^{j}_{M})^2$ and 
 $(\lt^{ij})^2 = (\l^{i}_{\mu} - \l^{j}_{\mu})^2 + (\l^{i}_{m} + \l^{j}_{m})^2$, 
 which correspond to each distance between the eigenvalues. 
The former $(\l^{ij})^2$ is the second power of distance between two spacetime points 
 and the latter $(\lt^{ij})^2$ is that between a spacetime point and a mirror image point. 
The interaction part is 
\begin{align}
 S_{int} 
  &= - \frac{1}{g^2} \tr [\l_{M} , \vt_{M}] [\vt_{M} , \vt_{N}] 
     - \frac{1}{4g^2} \tr [\vt_{M} , \vt_{N}]^2 
     - \frac{1}{g^2} \tr [x'_{M} , \vt_{N}] [\l_{M} , \vt_{N}] \cr 
  &~~~ + \frac{1}{g^2} \tr [x'_{M} , \vt_{M}] [\l_{N} , \vt_{N}]  
     - \frac{1}{g^2} \tr [\l_{M} , b] [\vt_{M} , c] 
     - \frac{1}{g^2} \tr [\l_{M} , b] [x'_{M} , c].
\end{align}
The action (\ref{action:b}) is also written as 
\begin{align}
 S_b 
 &= -\frac{1}{2g^2} \tr [A_{M} , A_{N}]^2 + \cdots. 
\end{align}
The first term is the bosonic action of the $D=D_{ad} + D_{as}=10$ IIB matrix model and
 the remaining terms appear by the effect of orientifolding, which is our interest. 
The two-loop effective action, therefore, is written as 
\begin{align}
 W_2(\l) =  g^2 W_2^{\text{IIB}}(\l) + g^2 W'_2 (\l). 
\end{align}
Here the first term is the two-loop effective action for the IIB matrix model \cite{HNT}, 
\be
 W_2^{\text{IIB}}(\l) = 
  \frac{1}{2}(D-2)^2 I_1 - \frac{1}{2}D(3D-7)I_2 - 2(D-2) I_3, 
\ee
where 
\begin{align}
 &I_1 \equiv \sum_{i,j,k, j \neq k} \frac{1}{(\l^{ij})^2(\l^{ik})^2}, \cr
 &I_2 \equiv \sum_{i,j} \frac{1}{(\l^{ij})^4}, \cr
 &I_3 \equiv \sum_{i,j,k, j \neq k} \frac{1}{(\l^{ij})^2 (\l^{ik})^2} 
 	\frac{\l_{M}^{ij}}{\l_{M}^{ik}}. 
\end{align} 

In what follows we calculate concretely the remaining part 
 of the two-loop effective action. 
In this section, 
we pay attention to the action constituted by only the matrices of the adjoint representation, 
$M_{\mu}$, $N_{\mu}$ and the ghosts.    
The part including the matrices of the antisymmetric representation is 
 calculated in Appendix A.
The action that we need now is 
\begin{align}
 S^{ad}_{int} = &\sum_{i,j} \sum_{\mu,\nu} 
  \left\{ 
    -\frac{1}{g^2} \left[ 2 \l_{\mu}^{ij} M_{\nu}^{ij} E^{ji}_{\mu\nu} 
    -\lt_{\mu}^{ij} (N_{\nu}^{ij} F_{\mu\nu}^{ij*} + N_{\nu}^{ij*} F_{\mu\nu}^{ij} ) 
    \right] \right. %\cr &~~~~~~
    -\frac{1}{2g^2} 
    \left[ E_{\mu\nu}^{ij} E_{\mu\nu}^{ji} - F_{\mu\nu}^{ij} F_{\mu\nu}^{ij*} \right] \cr 
    &~~~~~~ \left.-\frac{1}{g^2} \left[ 2\l^{ij}_{\mu} b_{(1)}^{ij} E_{(c) \mu}^{ji}
     - \lt^{ij}_{\mu} (b_{(2)}^{ij} F_{(c) \mu}^{ij*} + b_{(2)}^{ij*} F_{(c)\mu}^{ij}) \right]
    \right\} \cr
    & -\frac{4}{g^2} \left\{ \sum_{ij,k \neq i} \left[
    \frac{\l_{\mu}^{ij}}{\l_{\mu}^{ik}} M_{\mu}^{ik} M_{\mu}^{ki} M_{\nu}^{ij} M_{\nu}^{ji} 
    + \frac{\lt_{\mu}^{ij}}{\l_{\mu}^{ik}} M_{\mu}^{ik} M_{\mu}^{ki} N_{\nu}^{ij} N_{\nu}^{ji*}
    \right]
    \right. \cr
    &~~~~~~ \left. 
    + \sum_{ijk} \left[
    \frac{\lt_{\mu}^{ij}}{\lt_{\mu}^{ik}} N_{\mu}^{ik}N_{\mu}^{ki*}N_{\nu}^{ij}N_{\nu}^{ji*}
    + \frac{\l_{\mu}^{ij}}{\lt_{\mu}^{ik}} N_{\mu}^{ik}N_{\mu}^{ki*}M_{\nu}^{ij}M_{\nu}^{ji}
    \right]
    \right\} \cr 
    & +\frac{2}{g^2} \left\{ \sum_{ij,k \neq i} \left[
    \frac{\l_{\nu}^{ij}}{\l_{\mu}^{ik}} M_{\mu}^{ik} M_{\mu}^{ki} 
     (M_{\mu}^{ij}M_{\nu}^{ji} + M_{\mu}^{ji}M_{\nu}^{ij} )
    + \frac{\lt_{\nu}^{ij}}{\l_{\mu}^{ik}} M_{\mu}^{ik} M_{\mu}^{ki} 
     (N_{\mu}^{ij}N_{\nu}^{ij*} + N_{\mu}^{ij*}N_{\nu}^{ij} )
    \right] 
    \right. \cr &~~~~~~ \left. 
    + \sum_{ijk} \left[
    \frac{\lt_{\nu}^{ij}}{\lt_{\mu}^{ik}} N_{\mu}^{ik} N_{\mu}^{ik*} 
     (N_{\mu}^{ij}N_{\nu}^{ij*} + N_{\mu}^{ij*}N_{\nu}^{ij} )
    + \frac{\l_{\nu}^{ij}}{\lt_{\mu}^{ik}} M_{\mu}^{ik} M_{\mu}^{ki} 
     (N_{\mu}^{ij}N_{\nu}^{ij*} + N_{\mu}^{ij*}N_{\nu}^{ij} )
    \right] \right\} \cr
    & - \frac{2}{g^2} \left\{ \sum_{ij,k \neq i} \left[
    \frac{\l_{\mu}^{ij}}{\l_{\mu}^{ik}} M_{\mu}^{ik} M_{\mu}^{ki} 
     (b_{(1)}^{ij}c_{(1)}^{ji} + b_{(1)}^{ji}c_{(1)}^{ij})
    + \frac{\lt_{\mu}^{ij}}{\l_{\mu}^{ik}} M_{\mu}^{ik} M_{\mu}^{ki} 
     (b_{(2)}^{ij}c_{(2)}^{ij*} + b_{(2)}^{ij*}c_{(2)}^{ij} )
    \right] 
    \right. \cr &~~~~~~ \left. 
    + \sum_{ijk} \left[
    \frac{\lt_{\mu}^{ij}}{\lt_{\mu}^{ik}} N_{\mu}^{ik} N_{\mu}^{ik*} 
     (b_{(2)}^{ij}c_{(2)}^{ij*} + b_{(2)}^{ij*}c_{(2)}^{ij} )
    + \frac{\l_{\mu}^{ij}}{\lt_{\mu}^{ik}} N_{\mu}^{ik} N_{\mu}^{ik*} 
     (b_{(1)}^{ij}c_{(1)}^{ji} + b_{(1)}^{ji}c_{(1)}^{ji} ) 
    \right] \right\}, 
\end{align}
where 
\begin{align}
E_{\mu\nu}^{ij} &= 
 [M_{\mu},M_{\nu}]^{ij} + (N_{\mu}N_{\nu}^{*})^{ij} - (N_{\nu}N_{\mu}^{*})^{ij}, \\
F_{\mu\nu}^{ij} &= 
 [(M_{\mu} N_{\nu})^{ij} + (M_{\mu}n_{\nu})^{ji}] 
 - [(M_{\nu} N_{\mu})^{ij} + (M_{\nu}N_{\mu})^{ji}], \\
E_{(c)\mu}^{ij} &= 
 [M_{\mu},c_{(1)}]^{ij} + (N_{\mu}c_{(2)}^{*})^{ij} - (c_{(2)}N_{\mu}^{*})^{ij}, \\
F_{(c)\mu}^{ij} &= 
 [(M_{\mu} c_{(2)})^{ij} + (M_{\mu}c_{(2)})^{ji}] 
 - [(c_{(1)} N_{\mu})^{ij} + (c_{(1)}N_{\mu})^{ji}].
\end{align}

%--------------------------FIGURE-----------------------------
\begin{figure}[h]
\begin{center}
\includegraphics[scale=0.6]{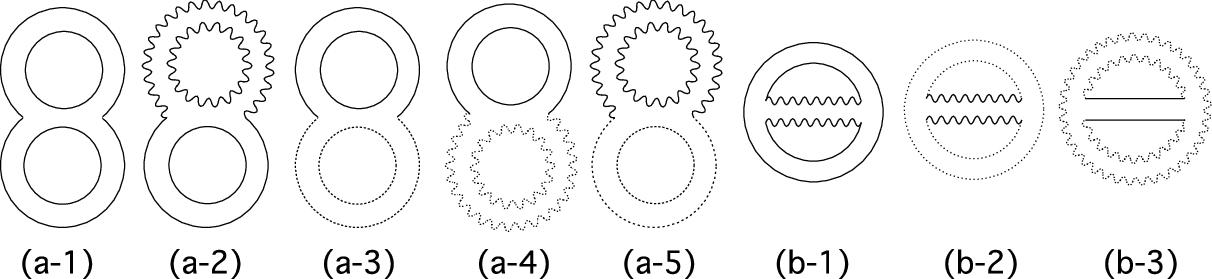}
\caption{two-loop planar diagrams.
}
%\label{fig:diag}
\end{center}
\end{figure}
%--------------------------FIGURE-----------------------------

The two-loop planar diagrams are shown in Figure 2. 
The solid line and the wavy line 
represent the propagator of $N_{\mu}^{ij}$ and $M_{\mu}^{ij}$, respectively. 
Similarly, the dashed solid line and wavy line correspond to the ghosts 
$b_{(2)}, c_{(2)}$ and $b_{(1)}, c_{(1)}$.
The diagram (a-1) in Figure 2 %constructed by the matrix $N_{\mu}$ 
 is evaluated as   
\begin{align}
 \text{(a-1)} &= 
 \left\langle 
  \sum_{i,j,k,l} \sum_{\mu,\nu} 
  (N_{\mu}^{ik} N_{\nu}^{kj*} N_{\mu}^{jl} N_{\nu}^{li*} 
  - N_{\mu}^{ik*} N_{\mu}^{kj} N_{\nu}^{jl*} N_{\nu}^{li}) \right. \cr
  &\left. ~~~~~+ \sum_{i,j,k} \sum_{\mu,\nu} 
  4 \frac{\lt_{\mu}^{ij}}{\lt_{\mu}^{ik}} N_{\mu}^{ik} N_{\mu}^{ki*} N_{\nu}^{ij} N_{\nu}^{ji*} 
  - \sum_{i,j,k} \sum_{\mu,\nu} 2 \frac{\lt_{\nu}^{ij}}{\lt_{\mu}^{ik}} 
  N_{\mu}^{ik} N_{\nu}^{ik*} (N_{\mu}^{ij} N_{\nu}^{ij*} + N_{\mu}^{ij*} N_{\nu}^{ij})
  \right\rangle \cr
  &=  - \frac{1}{2} D_{ad}(D_{ad}-1) {J}^{+}_1 
  + \frac{3}{2} D_{ad}(D_{ad}-1) {J}^{+}_2 
  + 2 (D_{ad}-1) {J}_3, 
\label{mmmm}
\end{align}
where 
\begin{align}
{J}^{\pm}_1 &= \sum_{i,j,k \neq j} \frac{1}{2} \frac{1}{(\lt^{ij})^2 (\lt^{ik})^2}
	\left( 
	1 % + \frac{\d^{ij}\d^{ik}}{(\lt^{ii})^4} 
        \pm \d^{ij} 
        \pm \d^{ik}
        \right),
        \cr
{J}^{\pm}_2 &= \sum_{i,j} \frac{1}{2} \frac{1}{(\lt^{ij})^4}
	 \left( 
	1 + \d^{ij} \pm 2 \d^{ij}
        % + \frac{2}{(\lt^{ii})^2 (\lt^{ij})^2}
	\right), \cr
{J}_3 &= \sum_{i,j,k \neq j} \sum_{\mu} \frac{1}{2} 
        \frac{\lt^{ij}_{\mu}}{\lt^{ik}_{\mu}} 
        \frac{1}{(\lt^{ij})^2 (\lt^{ik})^2} 
        \left(
        1 %+ \d^{ij} \d^{ik} 
        + \d^{ik} + \d^{ij}  
        \right). 
\end{align}
Similarly, the diagrams (a-2) $\sim$ (b-3)  are evaluated as 
\begin{align}
 \text{(a-2)} &= - 2 D_{ad}(D_{ad}-1) {K}^{(1)}_1 + 4 (D_{ad}-1) {K}^{(1)}_3 %\\
 		+ 4 (D_{ad}-1) \tilde{K}^{(1)}_3, \\
 \text{(a-3)} &= -2 (D_{ad} {J}^{+}_2 + {J}_3), \\
 \text{(a-4)} &= -4 {K}^{(1)}_3, \\ 
 \text{(a-5)} &= -4 \tilde{K}^{(1)}_3, \\
 \text{(b-1)} &= 2 (D_{ad} - 1) {L}^{(1)}_1 + 2 (D_{ad} - 1) {L}^{(1)}_2
 		+ 4 (D_{ad} - 1) \tilde{L}^{(1)}_2,\\
 \text{(b-2)} &= - 2 {L}^{(1)}_2, \\
 \text{(b-3)} &= {L}_3,
\end{align}
where
\begin{align}
{K}^{\pm}_1 &= \sum_{i,j,k \neq j} \frac{1}{2} 
	\frac{1}{(\l^{jk})^2 (\lt^{ij})^2}
	\left( 
	 1 + \d^{ik} \pm \d^{ij}
        \right), \cr
{K}^{(1)}_3 &= \sum_{i,j,k \neq i} \sum_{\mu} \frac{1}{2} 
	%\left\{ 
        \frac{\l_{\mu}^{ik}}{\lt_{\mu}^{ij}} 
        %+ \frac{\lt_{\mu}^{ij}}{\l_{\mu}^{ik}} \right\}
	\frac{1}{(\l^{ik})^2(\lt^{ij})^2} 
        \left( 
	 1 + \d^{ij}
        \right), \cr
\tilde{K}^{(1)}_3 &= \sum_{i,j,k \neq i} \sum_{\mu} \frac{1}{2} 
	\frac{\lt_{\mu}^{ij}}{\l_{\mu}^{ik}} 
	\frac{1}{(\l^{ik})^2(\lt^{ij})^2} 
        \left( 
	 1 + \d^{ij}
	\right), \cr
%I'''_3 &= \sum_{i,j\neq i} \frac{(\l^{ij}_{\mu})^2}{(\l^{ij})^2} 
%	\left[ \frac{1}{(\lt^{jj})^4} + \frac{1}{(\lt^{ij})^4} 
%        	+ \sum_k \frac{1}{(\lt^{jk})^4} \right]
{L}^{(1)}_1 &= \sum_{i,j,k\neq i} \sum_{\mu} \frac{1}{2} 
	\frac{(\l^{ik}_{\mu})^2}{(\l^{ik})^2} \frac{1}{(\lt^{ij})^2 (\lt^{kj})^2}
        \left(
	1 + \d^{ij} + \d^{jk}
        \right), \cr
{L}^{(1)}_2 &= \sum_{i,j,k\neq i} \sum_{\mu} \frac{1}{2} 
        \frac{\lt^{ij}_{\mu} \lt^{kj}_{\mu}}{(\l^{ik})^2 (\lt^{ij})^2 (\lt^{jk})^2}
        (1 + \d^{ij} + \d^{jk}), \cr
\tilde{L}^{(1)}_2 &= \sum_{i,j,k\neq i} \sum_{\mu} \frac{1}{2} 
        \frac{\lt^{ij}_{\mu} \lt^{kj}_{\mu}}{(\l^{ik})^2 (\lt^{ij})^2 (\lt^{jk})^2}
        (1 + \d^{ij} + \d^{jk}), \cr
{L}_3 &= \sum_{i,j} \sum_{\mu} \frac{1}{4}
	\frac{\l^{ij}_{\mu} \lt^{ij}_{\mu}}{(\l^{ij})^2 (\lt^{ij})^2 (\lt^{jj})^2}
        (1 + \d^{ij} + \d^{jk}). 
\end{align}
Since the two-loop effective action is given by summing up all corrections 
 and flipping the sign, 
we obtain the adjoint part, 
\begin{align}
 W^{ad}_2(\l) = 
  &\frac{1}{2} D_{ad} (D_{ad}-1) J_1^{+} - \frac{1}{2}D_{ad}(3D_{ad}-7) J^{+}_2
  		- 2 (D_{ad}-2) J_3 \cr
                &+ 2 D_{ad} (D_{ad}-1) K^{+}_1 - 4 (D_{ad}-2) K^{(1)}_3
                - 4 (D_{ad}-2) \tilde{K}^{(1)}_3 \cr 
                &- 2 (D_{ad}-1) L^{(1)}_1 - 2(D_{ad} - 2) L^{(1)}_2
                - 2(D_{ad}-1) \tilde{L}^{(1)}_2 - L_3. 
\end{align}
When the result calculated in the Appendix A is included,
we can get the orientifolding effect part of the two-loop effective action, 
\be
 W'(\l) = W^{ad}_2(\l) + W^{as}_2(\l) + W^{int}_2(\l), 
\label{2-loop}
\ee
where $W^{as}_2(\l)$ and $W^{int}_2(\l)$ are given 
 in equation (\ref{2-loop;as}) and (\ref{2-loop;other}). 

%The two-loop effective action is
%\begin{align}
% W_2(\l) = 
%\end{align}

%%%%%%%%%%%%%%%%%%%%%%%%%%%%%%%%%%%%%%%%%%%%%%%%%%%%%%%%%%%%%%%%%%%%%%%%%%%%%%%%%%%%%%
%%%%%%%%%%%%%%%%%%%%%%%%%%%%%%%%%%%%%%%%%%%%%%%%%%%%%%%%%%%%%%%%%%%%%%%%%%%%%%%%%%%%%%
\section{Discussion}
%%%%%%%%%%%%%%%%%%%%%%%%%%%%%%%%%%%%%%%%%%%%%%%%%%%%%%%%%%%%%%%%%%%%%%%%%%%%%%%%%%%%%%

In the last section, the two-loop effective action have been calculated. 
First, according to the procedure used in the paper \cite{HNT}, 
 in order to discuss the extent of the spacetime described by the USp matrix model, 
 we estimate the order of magnitude of the two-loop correction. 
Then, we can determine the upper bound of the distance between two eigenvalues 
 in the USp matrix model. % following the procedure used in the paper \cite{HNT}.
The order of magnitude of the one-loop correction is   
\begin{align}
(\text{one-loop}) \sim  O(N^2), 
\end{align}
and that of the two-loop corrections which is obtained in the last section is 
\begin{align}
 (\text{two-loop}) \sim  O \left( N^2 \frac{g^2 N}{R^4} \right). 
\end{align}
where $R$ is the expectation value of the meanvalue of the distance 
 between the spacetime points.   
The two-loop corrections can be neglected for 
 $R > \sqrt{g} N^{\frac{1}{4}}$. 
Thus it is enough that we think of only the attraction 
 by the one-loop corrections in this region. 
As there are mirror image points with respect to  
 plane spanned by the antisymmetric directions, 
 the points are attracted to this $D_{as}$-dimensional plane. 
The spacetime points, therefore, are restricted in the region 
 whose distance from the antisymmetric plane in Figure 3 within $\sqrt{g} N^{\frac{1}{4}}$. 

%--------------------------FIGURE-----------------------------
\begin{figure}[h]
\begin{center}
\includegraphics[scale=0.55]{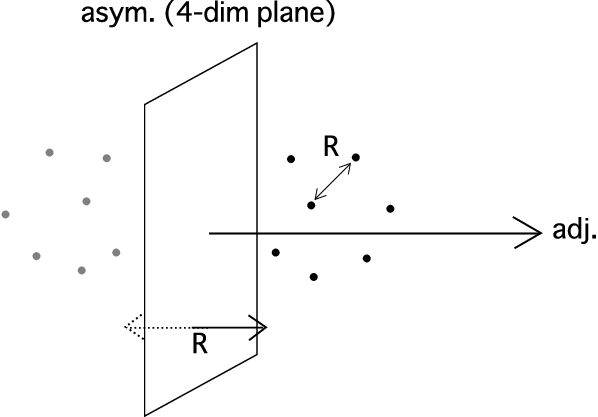}
\caption{
The points are attracted to the plane 
 spanned by the antisymmetric directions due to dominant one-loop effect 
 when the distance of the spacetime point from the plane is large. 
}
\label{fig:diag}
\end{center}
\end{figure}
%--------------------------FIGURE-----------------------------

We must take supersymmetry into account in this discussion. 
However although supersymmetry may adjust the two-loop corrections, 
 the above obtained upper bound is certain. 
In addition, since supersymmetry demands $D_{ad}=6$ and $D_{as}=4$, 
 the antisymmetric plane is four-dimensional. 

On the other hand, for each spacetime point, there is always a mirror image point 
 with respect to the plane spanned by the four antisymmetric directions
 by construction.  
In \cite{IY2} we found that two-body interaction between a spacetime point 
 and its mirror image point is described by the six-dimensional 
 SU(2) matrix model. 
In fact, in the two-loop effective action (\ref{2-loop}), 
 the interaction term between the point $i$ and its mirror image point $\bar{i}$,
 which contain the factor $(\lt^{ii})^4 = (2 \l_{\mu}^i)^4$ is written by 
\be
   W'(\l)|_{i\text{-}\bar{i} ~ \text{interaction}} = 
   - \frac{1}{2} D_{ad}(3 D_{ad} - 7) \frac{1}{(2\l^{i})^4}, 
\ee
which is the same as the interaction term between the eigenvalues 
 $\l^i_{\mu}$ and $- \l^i_{\mu}$ in $D_{ad}$-dimensional SU(2) matrix model. 
The number of the coordinates belonging to the antisymmetric representation 
 $D_{as}$ is not related to this interaction. 
In the SU(2) model, the expectation value of the distance between these two eigenvalues 
 is the quantity of the order of Planck length \cite{ST}.
This result suggests that the eigenvalues distribute 
 in a tubular region around the four directions of antisymmetric representation.

\vspace{5mm}
%%%%%%%%%%%%%%%%%%%%%%%%%%%%%%%%%%%%%%%%%%%%%%%%%%%%%%%%%%%%%%%%%%%%%%%%%%%%%%%%%%%%%%%%

{\bf{Acknowledgements}}\\
We would like to thank H. Itoyama for useful discussions on this paper. 
This work is supported %in part 
 by the Grant-in-Aid for Scientific Research (2054278).
\vspace{3mm}

%%%%%%%%%%%%%%%%%%%%%%%%%%%%%%%%%%%%%%%%%%%%%%%%%%%%%%%%%%%%%%%%%%%%%%%%%%%%%%%%%%%%%%%%
\appendix

\section{two-loop corrections}

In this appendix, we calculate the remaining effective action. 
The action constituted by the matrices of the antisymmetric representation only is 
\begin{align}
 S_{as}^{\text{int}} = &\sum_{i,j} \sum_{m,n} 
  \left\{ 
    -\frac{1}{g^2} \left[ 2 \l_{m}^{ij} A_{n}^{ij} E^{ji}_{mn} 
    -\l_{m}^{ij} (B_{n}^{ij} F_{mn}^{ij*} + B_{n}^{ij*} F_{mn}^{ij} ) 
    \right] \right. %\cr &~~~~~~
    -\frac{1}{2g^2} 
    \left[ E_{mn}^{ij} E_{mn}^{ji} - F_{mn}^{ij} F_{mn}^{ij*} \right] \cr 
    &~~~~~~ \left. -\frac{1}{g^2} \left[ 2\l^{ij}_{m} b_{(1)}^{ij} E_{(c) m}^{ji}
     + \l^{ij}_{m} (b_{(2)}^{ij} F_{(c) m}^{ij*} + b_{(2)}^{ij*} F_{(c)m}^{ij}) \right]
    \right\} \cr
    & -\frac{4}{g^2} \left\{ \sum_{ij,k \neq i} \left[
    \frac{\l_{m}^{ij}}{\l_{m}^{ik}} A_{m}^{ik} A_{m}^{ki} A_{n}^{ij} A_{n}^{ji} 
    - \frac{\l_{m}^{ij}}{\l_{m}^{ik}} A_{m}^{ik} A_{m}^{ki} B_{n}^{ij} B_{n}^{ji*}
    \right]
    \right. \cr
    &~~~~~~ \left. 
    + \sum_{ij,k \neq i} \left[
    \frac{\l_{m}^{ij}}{\l_{m}^{ik}} B_{m}^{ik}B_{m}^{ki*}B_{n}^{ij}B_{n}^{ji*}
    - \frac{\l_{m}^{ij}}{\l_{m}^{ik}} B_{m}^{ik}B_{m}^{ki*}A_{n}^{ij}A_{n}^{ji}
    \right]
    \right\} \cr 
    & +\frac{2}{g^2} \left\{ \sum_{ij,k \neq i} \left[
    \frac{\l_{n}^{ij}}{\l_{m}^{ik}} A_{m}^{ik} A_{m}^{ki} 
     (A_{m}^{ij}A_{n}^{ji} + A_{m}^{ji}A_{n}^{ij} )
    + \frac{\l_{n}^{ij}}{\l_{m}^{ik}} A_{m}^{ik} A_{m}^{ki} 
     (B_{m}^{ij}B_{n}^{ij*} + B_{m}^{ij*}B_{n}^{ij} )
    \right] 
    \right. \cr &~~~~~~ \left. 
    + \sum_{ijk} \left[
    \frac{\l_{n}^{ij}}{\l_{m}^{ik}} B_{m}^{ik} B_{m}^{ik*} 
     (B_{m}^{ij}B_{n}^{ij*} + B_{m}^{ij*}B_{n}^{ij} )
    + \frac{\l_{n}^{ij}}{\l_{m}^{ik}} B_{m}^{ik} B_{m}^{ik*} 
     (A_{m}^{ij}A_{n}^{ji} + A_{m}^{ij}A_{n}^{ji} )
    \right] \right\} \cr
    & - \frac{2}{g^2} \left\{ \sum_{ij,k \neq i} \left[
    \frac{\l_{m}^{ij}}{\l_{m}^{ik}} A_{m}^{ik} A_{m}^{ki} 
     (b_{(1)}^{ij}c_{(1)}^{ji} + b_{(1)}^{ji}c_{(1)}^{ij})
    + \frac{\l_{m}^{ij}}{\l_{m}^{ik}} A_{m}^{ik} A_{m}^{ki} 
     (b_{(2)}^{ij}c_{(2)}^{ij*} + b_{(2)}^{ij*}c_{(2)}^{ij} )
    \right] 
    \right. \cr &~~~~~~ \left. 
    + \sum_{ijk} \left[
    \frac{\l_{m}^{ij}}{\l_{m}^{ik}} B_{m}^{ik} B_{m}^{ik*} 
     (b_{(2)}^{ij}c_{(2)}^{ij*} + b_{(2)}^{ij*}c_{(2)}^{ij} )
    + \frac{\l_{m}^{ij}}{\l_{m}^{ik}} B_{m}^{ik} B_{m}^{ik*} 
     (b_{(1)}^{ij}c_{(1)}^{ji} + b_{(1)}^{ji}c_{(1)}^{ji}) 
    \right] \right\} 
\end{align}
where 
\begin{align}
E_{mn}^{ij} &= 
 [A_{m},A_{n}]^{ij} - (B_{m}B_{n}^{*})^{ij} + (B_{n}B_{m}^{*})^{ij} \\
F_{mn}^{ij} &= 
 [(A_{m} B_{n})^{ij} + (A_{m}n_{n})^{ji}] 
 - [(A_{n} B_{m})^{ij} + (A_{n}B_{m})^{ji}] \\
E_{(c)m}^{ij} &= 
 [A_{m},c_{(1)}]^{ij} + (B_{m}c_{(2)}^{*})^{ij} - (c_{(2)}B_{m}^{*})^{ij} \\
F_{(c)m}^{ij} &= 
 [(A_{m} c_{(2)})^{ij} + (A_{m}c_{(2)})^{ji}] 
 - [(c_{(1)} B_{m})^{ij} + (c_{(1)}N_{m})^{ji}].
\end{align}
The two-loop planar diagrams are the same as Figure 2. 
However, the solid line and the wavy line 
represent the propagator of $B_{m}^{ij}$ and $A_{m}^{ij}$, respectively 
in this case.  
We can obtain the following results from each diagram in Figure 2; 
\begin{align}
\text{(a-1)} &= -\frac{1}{2} D_{as} (D_{as} - 1) {J}_1^{-} 
 + \frac{3}{2} D_{as} (D_{as} - 1) {J}_2^{-} 
 + 2 (D_{as} - 1) {J}'_3 \cr
\text{(a-2)} &= - 2 D_{as}(D_{as}-1) {K}^{-}_1 + 4 (D_{as}-1) {K}^{(2)}_3 
 		+ 4 (D_{as}-1) {K}'^{(2)}_3 \cr
 \text{(a-3)} &= -2 (D_{as} {J}^{-}_2 + {J}'_3) \cr
 \text{(a-4)} &= -4 {K}^{(2)}_3 \cr 
 \text{(a-5)} &= -4 \tilde{K}^{(2)}_3 \cr 
 \text{(b-1)} &= 6 (D_{as} - 1) {L}^{(2)}_1 + 2 (D_{as} - 1) {L}^{(2)}_2 \cr 
 \text{(b-2)} &= - 2 {L}^{(2)}_2\cr 
 \text{(b-3)} &= {L}^{(2)}_2
\end{align}
where 
\begin{align}
 {J}'_3 &= \sum_{i,j,k \neq j} \sum_{m} \frac{1}{2} 
        \frac{\l^{ij}_{m}}{\l^{ik}_{m}} 
        \frac{1}{(\lt^{ij})^2 (\lt^{ik})^2}, \cr
 {K}^{(2)}_3 &= \sum_{j,k,i \neq j,k} \sum_{m} \frac{1}{2} 
	%\left\{ 
        \frac{\l_{m}^{ik}}{\l_{m}^{ij}} 
        %+ \frac{\lt_{\mu}^{ij}}{\l_{\mu}^{ik}} \right\}
	\frac{1}{(\l^{ik})^2(\lt^{ij})^2}, \cr
\tilde{K}^{(2)}_3 &= \sum_{i,j,k \neq i} \sum_{m} \frac{1}{2} 
	\frac{\l_{m}^{ij}}{\l_{m}^{ik}} 
	\frac{1}{(\l^{ik})^2(\lt^{ij})^2}, \cr
 {L}^{(2)}_1 &= \sum_{i,j,k\neq i} \sum_{m} \frac{1}{2} 
	\frac{(\l^{ik}_{m})^2}{(\l^{ik})^2} \frac{1}{(\lt^{ij})^2 (\lt^{kj})^2}
        \left(
	1 - \d^{ij} - \d^{jk}
        \right), \cr
{L}^{(2)}_2 &= \sum_{i,j,k\neq i} \sum_{m} \frac{1}{2} 
        \frac{\l^{ij}_{m} \l^{kj}_{m}}{(\l^{ik})^2 (\lt^{ij})^2 (\lt^{jk})^2}, 
\end{align}
Therefore we obtain 
\begin{align}
 W_2^{as}(\l) =  
 	&\frac{1}{2} D_{as} (D_{as}-1) J_1^{-} - \frac{1}{2}D_{as}(3D_{as}-7) J^{-}_2
  	- 2 (D_{as}-2) J'_3 \cr
        &+ 2 D_{as} (D_{ad}-1) K^{-}_1 - 4 (D_{as}-2) K^{(2)}_3
 	- 4 (D_{as}-2) K'^{(2)}_3 \cr 
        &- 6 (D_{as}-1) L^{(2)}_1 - (2 D_{as} - 3) L^{(3)}_2.
 \label{2-loop;as}
\end{align}
Finally, the remainder of the interaction terms are 
\begin{align}
 S'^{\text{int}} = &\sum_{i,j} \sum_{m,n} 
  \left\{ 
    - \frac{1}{g^2} \left[ 2 \l_{\mu}^{ij} A_{n}^{ij} E^{ji}_{\mu n} 
    + \lt_{\mu}^{ij} (B_{n}^{ij} F_{\mu n}^{ij*} + B_{n}^{ij*} F_{\mu n}^{ij} ) 
    \right] \right. %\cr &~~~~~~
    - \frac{1}{2g^2} 
    \left[ E_{\mu n}^{ij} E_{\mu n}^{ji} + F_{\mu n}^{ij} F_{\mu n}^{ij*} \right] \cr 
    &~~~~~~ \left.
    - \frac{1}{g^2} \left[ 2 \l_{m}^{ij} M_{\nu}^{ij} E^{ji}_{m \nu} 
    + \l_{m}^{ij} (N_{\nu}^{ij} F_{m \nu}^{ij*} + N_{\nu}^{ij*} F_{m \nu}^{ij} ) 
    \right]  %\cr &~~~~~~
    -\frac{1}{2g^2} 
    \left[ E_{m\nu}^{ij} E_{m\nu}^{ji} + F_{m\nu}^{ij} F_{m\nu}^{ij*} \right] \right\} \cr 
    %&~~~~~~ 
    %\left. -\frac{1}{g^2} \left[ 2\l^{ij}_{m} b_{(1)}^{ij} E_{(c) m}^{ji}
    % + \l^{ij}_{m} (b_{(2)}^{ij} F_{(c) m}^{ij*} + b_{(2)}^{ij*} F_{(c)m}^{ij}) \right]
    %\right\} \cr
    & -\frac{4}{g^2} \left\{ \sum_{i,j,k \neq i} \left[
    \frac{\l_{\mu}^{ij}}{\l_{\mu}^{ik}} M_{\mu}^{ik} M_{\mu}^{ki} A_{n}^{ij} A_{n}^{ji} 
    - \frac{\lt_{\mu}^{ij}}{\l_{\mu}^{ik}} M_{\mu}^{ik} M_{\mu}^{ki} B_{n}^{ij} B_{n}^{ji*}
    \right]
    \right. \cr
    &~~~~~~ \left. 
    - \sum_{i,j,k \neq i} \left[
    \frac{\lt_{\mu}^{ij}}{\lt_{\mu}^{ik}} N_{\mu}^{ik} N_{\mu}^{ki*} B_{n}^{ij} B_{n}^{ji*}
    - \frac{\l_{\mu}^{ij}}{\lt_{\mu}^{ik}} N_{\mu}^{ik} N_{\mu}^{ki*} A_{n}^{ij} A_{n}^{ji}
    \right]
    \right\} \cr 
    & -\frac{4}{g^2} \left\{ \sum_{i,j,k \neq i} \left[
    \frac{\l_{m}^{ij}}{\l_{m}^{ik}} A_{m}^{ik} A_{m}^{ki} M_{\nu}^{ij} M_{\nu}^{ji} 
    + \frac{\l_{m}^{ij}}{\l_{m}^{ik}} A_{m}^{ik} A_{m}^{ki} N_{\nu}^{ij} N_{\nu}^{ji*}
    \right]
    \right. \cr
    &~~~~~~ \left. 
    - \sum_{i,j,k \neq i} \left[
    \frac{\l_{m}^{ij}}{\l_{m}^{ik}} B_{m}^{ik} B_{m}^{ki*} N_{\nu}^{ij} N_{\nu}^{ji*}
    + \frac{\l_{m}^{ij}}{\l_{m}^{ik}} B_{m}^{ik} B_{m}^{ki*} M_{\nu}^{ij} M_{\nu}^{ji}
    \right]
    \right\} 
\end{align}
where 
\begin{align}
E_{\mu n}^{ij} &= 
 [M_{\mu},A_{n}]^{ij} - (N_{\mu} B_{n}^{*})^{ij} - (B_{n} N_{\mu}^{*})^{ij} \cr
E_{m \nu}^{ij} &= 
 [A_{m},M_{\nu}]^{ij} + (B_{m} N_{\nu}^{*})^{ij} + (N_{\nu} B_{m}^{*})^{ij} \cr
F_{\mu n}^{ij} &= 
 [(M_{\mu} B_{n})^{ij} - (M_{\mu} B_{n})^{ji}] 
 - [(A_{n} N_{\mu})^{ij} - (A_{n} N_{\mu})^{ji}] \cr
F_{m \nu}^{ij} &= 
 [(A_{m} N_{\nu})^{ij} - (A_{m} N_{\nu})^{ji}] 
 - [(N_{\nu} B_{m})^{ij} - (N_{\nu} B_{m})^{ji}]. 
\end{align}
From the action $S'^{int}$ we have the diagram (a-1), (a-2) and (b-1) 
 because the ghost terms are absent. 
The results are 
\begin{align}
\text{(a-1)} &= - D_{ad} D_{as} {J}_1 
 + 3 D_{ad} D_{as} {J}_2^{-} 
 + 2 D_{as} \tilde{J}_3 + 2 D_{ad} {J}'_3  \\
\text{(a-2)} &= 2 D_{ad} D_{as} {K}_1 
		+ 4 D_{as} ({K}^{(1)}_3 + {K}'^{(1)}_3) 
                + 4 D_{ad} ({K}^{(2)}_3 + \tilde{K}^{(2)}_3) \\
 \text{(b-1)} &= 4 D_{as} {L}'^{(1)}_1 + 4 D_{as} {L}'^{(2)}_1 
 		+2 D_{as} {L}'^{(1)}_2 + 2 D_{ad} {L}^{(2)}_2
\end{align}
where
\begin{align}
 {J}_1 &= \sum_{i,j,k \neq j} \frac{1}{2} \frac{1}{(\lt^{ij})^2 (\lt^{ik})^2}, \cr
 \tilde{J}_3 &= \sum_{i,j,k \neq j} \sum_{\mu} \frac{1}{2} 
        \frac{\lt^{ij}_{\mu}}{\lt^{ik}_{\mu}} 
        \frac{1}{(\lt^{ij})^2 (\lt^{ik})^2} 
        \left(
        1 %+ \d^{ij} \d^{ik} 
        + \d^{ik} - \d^{ij}  
        \right), \cr
 K_1 &= \sum_{i,j,k \neq j} \frac{1}{2} 
	\frac{1}{(\l^{jk})^2 (\lt^{ij})^2}
	\left( 
	 1 + \d^{ik}
        \right), \cr
 {K}'^{(1)}_3 &= \sum_{i,j,k \neq i} \sum_{\mu} \frac{1}{2} 
	%\left\{ 
        \frac{\l_{\mu}^{ik}}{\lt_{\mu}^{ij}} 
        %+ \frac{\lt_{\mu}^{ij}}{\l_{\mu}^{ik}} \right\}
	\frac{1}{(\l^{ik})^2(\lt^{ij})^2} 
        \left( 
	 1 - \d^{ij}
        \right), \cr 
{L}'^{(1)}_1 &= \sum_{i,j,k\neq i} \sum_{\mu} \frac{1}{2} 
	\frac{(\l^{ik}_{\mu})^2}{(\l^{ik})^2} \frac{1}{(\lt^{ij})^2 (\lt^{kj})^2}, \cr
{L}'^{(2)}_1 &= \sum_{i,j,k\neq i} \sum_{m} \frac{1}{2} 
	\frac{(\l^{ik}_{m})^2}{(\l^{ik})^2} \frac{1}{(\lt^{ij})^2 (\lt^{kj})^2}, \cr
{L}'^{(1)}_2 &= \sum_{i,j,k\neq i} \sum_{\mu} \frac{1}{2} 
        \frac{\lt^{ij}_{\mu} \lt^{kj}_{\mu}}{(\l^{ik})^2 (\lt^{ij})^2 (\lt^{jk})^2}
        (1 - \d^{ij} - \d^{jk}). 
\end{align}
Therefore 
\begin{align}
 W_2^{int} (\l) &=  D_{ad} D_{as} {J}_1 
 - 3 D_{ad} D_{as} {J}_2^{-} 
 - 2 D_{as} \tilde{J}_3 - 2 D_{ad} {J}'_3  \cr
 & -2 D_{ad} D_{as} {K}_1 
 - 4 D_{as} ({K}^{(1)}_3 + {K}'^{(1)}_3) 
 - 4 D_{ad} ({K}^{(2)}_3 + \tilde{K}^{(2)}_3) \cr
  & - 4 D_{as} {L}'^{(1)}_1 - 4 D_{as} {L}'^{(2)}_1 
  -2 D_{as} {L}'^{(1)}_2 - 2 D_{ad} {L}^{(2)}_2.
 \label{2-loop;other}
\end{align}

%%%%%%%%%%%%%%%%%%%%%%%%%%%%%%%%%%%%%%%%%%%%%%%%%%%%%%%%%%%%%%%%%%%%%%%%%%%%%%%%%%%%%%%%
%%%%%%%%%%%%%%%%%%%%%%%%%%%%%%%%%%%%%%%%%%%%%%%%%%%%%%%%%%%%%%%%%%%%%%%%%%%%%%%%%%%%%%%%

%%%%%%%%%%%%%%%%%%%%%%   END   %%%%%%%%%%%%%%%%%%%%%%%
\end{document}